\documentstyle[12pt]{article}
\begin{document}

\pagestyle{plain}
\setcounter{page}{1}
\newcounter{bean}
\baselineskip16pt

 \begin{titlepage}
 \begin{flushright}
SU-ITP 97/47 \\
hep-th/9711063
\end{flushright}

\vspace{7 mm}

\begin{center}
{\huge A note on discrete light cone quantization}

\vspace{5mm}

{\huge }
\end{center}
\vspace{10 mm}
\begin{center}
{\large
D. Bigatti$^\dagger$ and L. Susskind } \\
\vspace{3mm}
Stanford University

\vspace{7mm}

{\large Abstract} \\
\end{center}

\vspace{3mm}

In this brief note we would like to discuss, in a simple model
system, the
conditions under which the discrete light cone quantization
framework should be trusted as an approximation scheme, with
regard, in particular, to the size and mass of the system.
Specifically, we are going to discuss
``quark-antiquark'' bound states in 1+1 dim., for which a
natural size is provided by analogy with a ``two points and a
spring'' system, and show that the condition for obtaining a
reliable estimate is the same as the one derived in a recent
paper for black holes in matrix theory.
\noindent

\vspace{7mm}
\begin{flushleft}
November 1997

\vspace{8cm}
$\dagger$ On leave of absence from Universit\`a di Genova, Italy

\end{flushleft}
\end{titlepage}

\newpage
\renewcommand{\baselinestretch}{1.1}

The light cone frame \cite{0} is a choice of coordinates, which
for some problems can be extremely convenient, consisting in
keeping unmodified all but one
spatial dimensions (called transverse) and replacing the other
space
coordinate (longitudinal) and the time coordinate with
\begin{eqnarray} \displaystyle{
X^{\pm}= \frac{x^1\pm x^0}{\sqrt{2}}
} \end{eqnarray}
The $X^+$ is to be thought of as a new ``time'' coordinate
(meaning that
we codify a Cauchy problem by assigning initial data on a
surface
$X^+=$ constant); the hamiltonian which describes time
evolution
is required to have the form
\begin{eqnarray} \displaystyle{
H= \frac{p_i^2}{2p_-} + \frac{M^2}{2p_-}
} \end{eqnarray}
thus giving a natural identification of the light cone description
of our
(relativistic) system with another, which has non relativistic
energy and a term of potential energy which turns out to be
Galilean invariant and
proportional to $1/p_-$. We want, however, to
stress that light cone quantization allows to bypass the
complexity
and the subtleties of a quantum field theory by relating it to a
system described by a non relativistic type Schr\"odinger
equation, which makes
it (in the situations in which it can be handled) very powerful. A
variant
of the method is the so-called discrete light cone quantization,
which
consists in a compactification in the lightlike (!) direction $X^-$
over a circle of length $2\pi R$, thus yielding a discrete spectrum
for $p_-$:
\begin{eqnarray} \displaystyle{
p_-=\frac{N}{R}
} \end{eqnarray}

Let us now turn to describe the 't Hooft model for
1+1 dimensional ``mesons'' \cite{1}. The theory contains
``quarks'' and non abelian
``gluons''; the gauge group is $U(n)$ with $n$ very large
(allowing perturbative treatment in $\frac{1}{n}$). One chooses
a description
in light cone coordinates (since the system is 1+1 dim.~the
transverse
coordinates are just absent), and, furthermore, the light cone
gauge condition
\begin{eqnarray} \label{gauge} \displaystyle{
A_-=0
} \end{eqnarray}
decouples the ghosts and linearizes the gauge field Lagrangian.
There is a subtlety concerning this gauge condition in the
compactified theory. It is well known that the compactified gauge
theories have gauge invariant global Wilson loop degrees of
freedom which in this case have the form
\begin{eqnarray} \displaystyle{
W= Tr \: P \: e^{i \oint A_- dx^-}
} \end{eqnarray}
Obviously if $W$ is not equal to $1$ there is an obstruction to
setting $A_-$ to zero. However it is possible to include $W$ in
the hamiltonian with a large coefficient in such a way as to
energetically force $W$ to $1$. This eliminates the Wilson loop
from the dynamics and allows the choice of gauge (\ref{gauge}).

The Lagrangian is
\begin{eqnarray} \displaystyle{
{\cal L} = - \frac{1}{2} \: {\mathrm {Tr}}(\partial_- A_+)^2 - \bar{q}^a
(\gamma_\mu \partial^\mu + m_{(a)} + g \gamma_- A_+) q^a
} \end{eqnarray}
(following the notation of \cite{1} it is convenient in the following
to define $g_0^2 \equiv g^2 n$)
and we remind that, being in 1+1 dim., the coupling constant $g$
has the dimension of a mass.

Since the $\gamma_+$,  $\gamma_-$ matrices are nilpotent,
the propagator of the quark is given by the expression
$\displaystyle{\frac{-ik_-}{m_{(a)}^2 +2k_+ k_- -i \epsilon}}$ ;
the gluon propagator is $\displaystyle{\frac{1}{k_-^2}}$ and
the quark-quark-gluon vertex is $-2 g$. (There are
no $\gamma$ matrices left). Considering the limit $n \rightarrow
\infty$, $g^2 n$ fixed,
the only contributions come from planar diagrams (we neglect
the possible subtleties which
may arise in presence of asymptotic states and lines which close
at infinity, since we will
be studying bound states) which have the shape of ladders with,
possibly, self energy
insertions for the ``quark'' propagator. It is possible to compute
the dressed propagator
for the ``quark'' (see \cite{1}):
 \begin{eqnarray} \label{5} \displaystyle{
\frac{\displaystyle{-\frac{i}{2}}}{\displaystyle{k_+ +  
\frac{\frac{m^2}{2}-
\frac{g_0^2}{2\pi}}{k_-} + \frac{g_0^2}{2 \pi \lambda} (sgn\:k_-) -i
\frac{\epsilon}{2k_-} }}
} \end{eqnarray}
where $\lambda$ is an infrared cutoff: $\lambda<|k_-|<\infty$.
(We won't care about the limit
$\lambda \rightarrow 0$ since the final results don't depend on
$\lambda$; let us notice,
however, that the pole of the propagator when $\lambda
\rightarrow 0$ is moved towards
very big $k_+$. This corresponds to existence of a ``confining''
potential).

It is worth pointing out that eq.~(\ref{5}) has exactly the right
form in order to
identify our system with an auxiliary non relativistic one. If we
consider the non relativistic amplitude for a particle which is in
$x$ for $\tau =0$ to be in $x'$ at $\tau = t$
\begin{eqnarray} \displaystyle{
\theta(t) \langle x| e^{i \, {\cal H}(p) \, t} | x' \rangle = F(x, x'; t)
} \end{eqnarray}
where ${\cal H}(p)$ does not depend on the time component of
$p_\mu$, and Fourier-transform the matrix elements, we get the
well-known Green function (if we like, the non relativistic
propagator
\begin{eqnarray} \label{7} \displaystyle{
\frac{1}{\omega - {\cal H}(p) \mp i \epsilon}
} \end{eqnarray}
If we keep in mind the interpretation of $k_+$ as the generator
of the $x^+$ (``time'')
translations, it is clear that the two first terms in the
denominators of (\ref{5}) and
(\ref{7}) should be identified ($\omega$ arises from the
$e^{i \omega t}$ in the Fourier
transform); as for ${\cal H}(p)$, in our context it should be
a function of the $k_-$
(``space translation'' generators) only. The imaginary part of
the pole is simply renamed,
$\epsilon ' \equiv \frac{\epsilon}{2 k_-}$; we don't expect any
subtleties in the limit
$k_- \rightarrow 0$, because the DLCQ keeps us away from the
danger and, besides, the final results obtained by means of an
i.~r.~cutoff do not contain it, so we trust our computations all
the same. Notice, incidentally, that within the non relativistic
interpretation the sum of
ladder diagrams becomes a computation in ordinary time
independent perturbation theory
for the Schr\"odinger equation eigenvalues/eigenfunctions in
quantum mechanics.

Reference \cite{1} computes the approximate eigenfunctions
and eigenvalues for the
two-particle bound state ``quark-antiquark'' in the regime of
small quark masses:
\begin{eqnarray} \label{8} \displaystyle{
\varphi^K (x) \simeq \sin (K \pi x)
} \end{eqnarray}
 \begin{eqnarray} \displaystyle{
M^2_{B.S.}= \pi g_0^2 K
} \end{eqnarray}
where $K \in {N}$, 
$M_{B.S.}$ mass of the $q \bar{q}$
bound state.

Does the criterion of reference \cite{2} for a good DLCQ
approximation, namely that the
extension of the system fits inside the compactification region,
apply to this system?

The linear extension of the bound state is
\begin{eqnarray} \displaystyle{
R_{B.S.} \sim \frac{M_{B.S.}}{g_0^2}
} \end{eqnarray}
To see it, consider a quark pair separated by a distance $L$.
Gauss' law tells us that there is
an electric flux between them and that its energy density is $g^2
n = g_0^2$. Thus the potential
energy is $g_0^2 L$ (a wierd ``linear spring'' in place of the usual
quadratic one).
In the rest frame of the bound state, where the total energy is
$M_{B.S.}$,
the energy is both kinetic and potential. The maximum size of the
system is reached (classically) when all the energy is potential,
that is when $L g_0^2 = M_{B.S.}$. \\
The condition $R_{B.S.} {\
\lower-1.2pt\vbox{\hbox{\rlap{$<$}\lower5pt\vbox{\hbox{$\sim$}
}}}\ } R_{COMP}$ gives immediately, by
multiplying by $p_-=M$ and using $Rp_-=N$,
\begin{eqnarray} \displaystyle{
N  {\
\lower-1.2pt\vbox{\hbox{\rlap{$>$}\lower5pt\vbox{\hbox{$\sim$}
}}}\ } M_{B.S.} R_{B.S.} = \frac{M_{B.S.}^2}{g_0^2}
\sim K
} \end{eqnarray}
which is exactly the form of the condition in \cite{2}.

It is clear at once that this is the correct criterion: if we want to
identify correctly
which is the eigenfunction in eq.~(\ref{8}) (that is, if we want to
resolve the quantum
number of the system), we need to make sure that we are
sampling at least as many Fourier
components as $K$, or the reconstruction will simply miss the
oscillations of the wave function and fail altogether.

\vspace{5mm}

We have shown in this article that in a very controllable context,
the criterion for the validity of the approximation introduced in
ref.~\cite{2} is satisfied, namely that a good approximation to the
properties of a system in DLCQ only requires $N$ to be of order
$MR$. We have chosen to examine the 't Hooft model because
of its simplicity but it should not be difficult to see that the
criterion applies to a wider set of examples. For example
replacing the fermionic quarks by bosons should not make
significant differences for states of high excitation.

\section*{Acknowledgments}
We would like to thank Igor Klebanov for crucial discussions.

\end{document}